\documentclass{elsart}

\usepackage{amsmath,amssymb,cite,lineno}
\usepackage{graphicx}
\usepackage{amsbsy}
\usepackage{ifpdf}

\ifpdf
\usepackage[%
  pdftitle={Instructions for use of the document class
    elsart},%
  pdfauthor={Simon Pepping},%
  pdfsubject={The preprint document class elsart},%
  pdfkeywords={instructions for use, elsart, document class},%
  pdfstartview=FitH,%
  bookmarks=true,%
  bookmarksopen=true,%
  breaklinks=true,%
  colorlinks=true,%
  linkcolor=blue,anchorcolor=blue,%
  citecolor=blue,filecolor=blue,%
  menucolor=blue,pagecolor=blue,%
  urlcolor=blue]{hyperref}
\else
\usepackage[%
  breaklinks=true,%
  colorlinks=true,%
  linkcolor=blue,anchorcolor=blue,%
  citecolor=blue,filecolor=blue,%
  menucolor=blue,pagecolor=blue,%
  urlcolor=blue]{hyperref}
\fi

\makeatletter
\def\elsartstyle{%
    \def\normalsize{\@setfontsize\normalsize\@xiipt{14.5}}
    \def\small{\@setfontsize\small\@xipt{13.6}}
    \let\footnotesize=\small
    \def\large{\@setfontsize\large\@xivpt{18}}
    \def\Large{\@setfontsize\Large\@xviipt{22}}
    \skip\@mpfootins = 18\p@ \@plus 2\p@
    \normalsize
}
\@ifundefined{square}{}{}
\makeatother

\newtheorem{theorem}{Theorem}[section]

\begin{document}

\bibliographystyle{elsart-num-sort}

\begin{frontmatter}

\title{Averaged dynamics of time-periodic advection diffusion equations in the
limit of small diffusivity}

\author{Tobias Sch{\"a}fer\corauthref{cor}}  
\corauth[cor]{Corresponding author.} 
\ead{tobias@math.csi.cuny.edu} 
\author{Andrew C. Poje,}
\author{Jesenko Vukadinovic} 

\address{Department of Mathematics, The College of Staten
        Island, City University of New York, New York}

\begin{abstract}
We study the effect of advection and small diffusion on 
passive tracers. The advecting velocity field is assumed
to have mean zero and to possess time-periodic stream lines.
Using a canonical transform
to action-angle variables followed by a Lie-transform,
we derive an averaged equation describing the effective motion of 
the tracers. An estimate for the time validity of the 
first-order approximation is established. For particular cases of 
a regularized vortical flow we present explicit formulas for the 
coefficients of the averaged equation both at first and at second order.
Numerical simulations indicate that the validity of the above 
first-order estimate extends to the second order.
\end{abstract}

\end{frontmatter}

\section{Introduction}

The characterization of the behavior of a passive but
diffusing scalar advected by a prescribed, smooth velocity field has been the subject of
intensive research going back at least as far as Batchelor \cite{Batchelor:1959}. 
Above and beyond the obvious practical importance in applications
ranging from micro-mixers to global climate dynamics, 'scalar turbulence' as exhibited by solutions of
the linear advection diffusion equation also provides an avenue for insight into the structure of the
Navier-Stokes equations \cite{Shraiman:2000}.

Despite the linearity of the governing equation, complete characterization of  
scalar solutions, especially the asymptotic decay of such solutions for vanishing
diffusivity,  continues to pose considerable difficulties even when restricted to
the case of planar flows.  Here we are concerned with the so-called Batchelor regime, 
where the spatial scale of the velocity ($l_v$) is assumed to be much larger than the diffusive length
scale ($l_\kappa$). For the scalar turbulence problem with periodic velocity fluctuations in time and space, 
rigorous homogenization techniques 
\cite{majda-kramer:1999,Bonn:2001,pavliotis:2002} can be applied to compute effective,
renormalized diffusivities on large time and space ($L \gg l_v$) scales assuming that
the initial distribution of the scalar field satisfies the scale separation $l_s \ll l_v$.
On the other hand, the situation for scalar fields with variations commensurate with both
the velocity field and the domain size, ($l_s \sim l_v \sim L \gg l_\kappa$) requires different
techniques. 

The present paper is motivated in part by the phenomenon of persistent
patterns, termed 'strange eigenmodes' \cite{pierrehumbert:1994}, that occur under the action
of periodic stirring. Such patterns, characterized by exponential decay of the scalar variance and 
 self-similar evolution of scalar density functions,
have been observed both numerically and experimentally {\cite{Voth2003}, \cite{Camassa:2007}}. Theoretical
predictions of the decay rates of the scalar and the connection between the observed eigenmodes and the phase
space of the underlying advection dynamics have been investigated {\cite{Antonsen:1996, Sukhatme:2002, Giona:2004, 
Fereday:2004, Popovych:2007}}, most often in the context of non-linear maps.

In the strange eigenmode regime, the decay of the scalar contrast can
be studied via Floquet theory. While well established for ODE's, the
existing theory for parabolic PDE's requires that the PDE satisfies a
restrictive spectral gap condition which the advection diffusion
equation fails for vanishing diffusivity (see \cite
{Kuchment:1993}). An alternate approach, shown by Chow \textit{et al.}
\cite{Chow:1994} for one dimensional parabolic equations, is to prove
the existence of an inertial manifold for the system and then apply
ordinary Floquet theory to the inertial form.  Our goal here is less
ambitious, but a potential first step in this direction. Following
Krol \cite{krol:1991}, we propose a formal averaging procedure for the
advection-diffusion equation when the velocity field is has zero mean
and possesses time-periodic stream lines. The approach is
perturbative, making explicit use of the disparity of time-scales
between the advective and diffusive operators.  Advection fields of
the form $u=u(\xi,t)=\bar u(\xi)f(t)$ guarantee that, in the case of
vanishing diffusion, the time-dependent system can be solved using
action-angle variables, and that tracer trajectories will be
time-periodic. The explicit solution in action-angle coordinates
allows the original equation to be written in a form suitable for
averaging.  By applying Lie transform techniques, we derive an
approximate averaged equation.  The fact that this resulting equation
- in contrast to the original problem - has time-independent
coefficients, facilitates the theoretical and numerical analysis
tremendously. The use of the Lie transform also allows relatively
straight-forward computation of higher order corrections.

The form of the paper is as follows. In section 2 we consider the transformation which places the
advection-diffusion equation in a form suitable for averaging. Lie transform 
techniques are used to average the equation in section 3. A proof of the convergence of solutions of the averaged equations 
to those of the original time dependent problem is given in section 4. An application to a specific
flow field, a periodically modulated, regularized vortex, along with numerical comparisons of the
solutions are given in sections 5 and 6.

\section{Action-angle variables}

We consider the advection diffusion equation in the following form
\begin{equation} \label{advec}
c_t +(u\cdot \nabla)\,c - \kappa \,\nabla^2\,c = 0.
\end{equation}
All functions depend on the spatial variable $\xi=(x,y)^T$ and the
time $t$. We look at (\ref{advec}) as an initial value problem,
assuming that
\begin{equation} \label{init_advec}
c(0,\xi)=c^{(s)}(\xi)
\end{equation}
is a known function. Incompressibility $\nabla\cdot u = 0$ implies
that the given velocity field $u$ is derived from a stream function $\Psi$
such that
\begin{equation}
u(\xi,t) = \nabla^{\perp}\Psi(\xi,t)
\end{equation}
where $\nabla^{\perp}\equiv (\partial_y,-\partial_x)$.

We assume that 
the stream function $\Psi$ is of the particular form
\begin{equation}
\Psi(\xi,t) = \bar \Psi(\xi)f(t)
\end{equation}
where the function $f$ is periodic in time with period $T$. For consistency in the averaging
which follows, we also require that $\langle f \rangle = \frac{1}{T} \int_{0}^{T} f(t) dt = 0$. 

A standard non-dimensionalization of (\ref{advec}) with velocity, length and time
scales given respectively by $(U,L,T)$ gives
\begin{equation} \label{scaleadvec}
c_t + \frac{1}{St} (u\cdot \nabla)\,c - \epsilon \,\nabla^2\,c = 0
\end{equation}
where the dimensionless groups are the Strouhal number, $St = U T/ L$, the ratio of the forcing
period to the advective time-scale, and $ \epsilon = T \kappa / L^2$, the ratio of the forcing
period to the diffusive time-scale. Throughout, we assume $St = {\mathcal{O}}(1)$ and $ \epsilon \ll 1$.
To clarify notation, we take $u = u/St $ throughout.

Our aim is to derive an equation of the form
\begin{equation}
\bar c_t + {\mathcal{L}}\,\bar c = 0
\end{equation}
with a time-independent local linear operator ${\mathcal{L}}$ that can
be used to construct approximative solutions to the initial value
problem given by (\ref{init_advec}) or (\ref{scaleadvec}),
in the limit of small
diffusivity $\kappa \ll 1$. Due to the fact that the leading order
evolution is given by the periodically varying advection operator, we
cannot apply averaging techniques directly. 
Instead, we seek a transformation to the Lagrangian frame which results in a new
equation where the coefficients of both the advective and diffusive operators are
zero-mean periodic functions of time. The transformed equation is then in a form suitable
for averaging.

For the restricted class of flow fields considered, the the proper transformation is simply to
action-angle coordinates of the underlying,  conservative advection equation. Introduce the function $F$ as
\begin{equation}
F(t) = \int_0^t f(t')\,dt'
\end{equation}
and write the tracer coordinate $\xi$ as a function of $F$. We then
obtain the autonomous Hamiltonian system
\begin{equation}
\frac{d\xi}{dF} = \nabla^{\perp}\bar\Psi(\xi)\,.
\end{equation}
Since this system integrable, there exists a canonical transformation
\cite{arnold:1989}
\begin{equation}
{\mathcal{C}}:(x,y)\rightarrow(J,\theta)
\end{equation}
such that the advection-diffusion equation (\ref{advec}) becomes 
\begin{equation} \label{advec_can}
c_t - f(t)\omega(J)c_{\theta} - \epsilon\left(\Gamma:\nabla\nabla + \delta\cdot\nabla\right)c = 0
\end{equation}
with a matrix $\Gamma=\Gamma(\theta,J)$ and a vector $\delta =
\delta(\theta,J)$, that are solely determined by the canonical
transformation ${\mathcal{C}}$. The advantage of the representation
(\ref{advec_can}) lies in the fact that the evolution of the
unperturbed problem is linear and given by
\begin{equation}
J = J_0, \qquad \theta = \theta_0-\omega(J)F(t)\,.
\end{equation}
Therefore, we can now use these stream lines as coordinates via the
transformation
\begin{equation}
c(t,J,\theta) = v\left(t,J,\tilde\theta=\theta-\omega(J)F(t)\right)
\end{equation}
and with the transformation rules
\begin{eqnarray*}
c_J &=& - \omega'Fv_{\tilde\theta}+v_J, \\
c_{JJ} &=& (\omega')^2F^2v_{\tilde\theta\tilde\theta}-2\omega'Fv_{\tilde\theta J}-\omega''Fv_{\tilde\theta}+v_{JJ}
\end{eqnarray*}
and the rescaling of time as $\tau=\epsilon t$ the equation for $v$
takes the form
\begin{equation} \label{advec_ready}
v_{\tau} = \left(\tilde \Gamma:\nabla\nabla + \tilde \delta\cdot\nabla\right)v\,.
\end{equation}
Since the only explicit time dependence in the coefficients
$\tilde\Gamma$ and $\tilde\delta$ is given in terms of the 
$T$-periodic function $F$, the equation (\ref{advec_ready})
is now suitable for averaging.

\section{Lie transform averaging}

In order to average (\ref{advec_ready}), we use a technique based on
Lie transforms first developed in the finite-dimensional
context \cite{nayfeh:1973} and then applied to cases involving an
infinite number of degrees of freedom
\cite{kodama:1985,gabitov-schaefer-etal:2000}.  The basic idea of a
Lie transform is to use a near identity transform of the type
\begin{equation} \label{lie_trafo}
v = \exp(\phi\cdot\nabla_L)V \;\;.
\end{equation}
The linear operator, $\phi\cdot\nabla_L$, is chosen to eliminate 
the explicit time dependence of the coefficient of an equation 
\begin{equation} \label{time_lie}
v_{\tau} = X(v,\tau)
\end{equation}
in order to  obtain an equation with time-independent coefficients of the
form
\begin{equation} \label{notime_lie}
V_{\tau} = Y(V)\,.
\end{equation}
Since the functionals $X$ and $Y$ depend on $v$ and all its  spatial derivatives,
the operator $\phi\cdot\nabla_L$ will be defined in our case as
\begin{equation}
\phi\cdot\nabla_L = \sum_{n,m}\phi_{nx,my}\frac{\partial^{(n+m)}}{\partial V_{nx,my}}
\end{equation}
where $\phi_{nx,my}=\partial^{(n+m)}\phi/\partial x^n \partial y^m$
and $V_{nx,my}=\partial^{(n+m)}V/\partial x^n \partial y^m$
respectively. The subscript at $\nabla_L$ distinguishes this operator
from the usual $\nabla$. The generating function $\phi$ also depends
on $V$ and all its derivatives. The idea is that the explicit time
dependence will be kept in $\phi$ rather than in the equation for $V$,
hence $\phi$ will also depend periodically on $\tau$. The
general transformation rule under which (\ref{time_lie}) transforms to
(\ref{notime_lie}) using (\ref{lie_trafo}) is
\cite{hasegawa-kodama:1995}
\begin{equation}
Y\cdot\nabla + \left(\frac{\partial}{\partial t}\mathrm{e}^{\phi\cdot\nabla_L}\right)
{\mathrm{e}}^{-\phi\cdot\nabla} = \mathrm{e}^{\phi\cdot\nabla_L}(X\cdot\nabla_L){\mathrm{e}}^{-\phi\cdot\nabla_L}
\end{equation}
Both terms can be conveniently expanded using the Campbell-Baker-Hausdorff formulae
\begin{equation}
\left(\frac{\partial}{\partial \tau}\mathrm{e}^{\phi\cdot\nabla_L}\right)
=\left(\phi_t + \frac{1}{2!}[\phi,\phi_t]_L+\frac{1}{3!}[\phi,[\phi,\phi_t]_L]_L+...\right)\cdot\nabla_L
\end{equation}
\begin{equation}
\mathrm{e}^{\phi\cdot\nabla_L}(X\cdot\nabla_L){\mathrm{e}}^{-\phi\cdot\nabla_L}
= \left(X+[\phi,X]_L+\frac{1}{2!}[\phi,[\phi,X]_L]_L+...\right)\cdot\nabla_L
\end{equation}
where the Lie commutator is defined through $[A,B]_L =
(A\cdot\nabla_L)B-(B\cdot\nabla_L)A$ and again the subscript
distinguishes the Lie commutator from the usual commutator.
We now expand both $Y$ and $\phi$ in a series in the small parameter
$\epsilon$ as
\begin{equation}
Y = Y_0+Y_1+...,\qquad \phi = \phi_1+\phi_2+...
\end{equation}
where ${\mathcal{O}}(Y_n)={\mathcal{O}}(\phi_n)=\epsilon^n$ and differentiation by
$\tau$ lowers the order of $\phi_n$ by one
\begin{equation}
{\mathcal{O}}(\partial \phi_n/\partial\tau) = \epsilon^{n-1}\,.
\end{equation}
The equation for $Y$ can then be solved order by order. At the
leading order, we find
\begin{equation}
Y_0 + \frac{\partial\phi_1}{\partial \tau} = X
\end{equation}
and averaging this equation yields directly $Y_0=\langle X \rangle$
due to the periodicity of $\phi_1$.

The transformed advection-diffusion equation (\ref{advec_ready}) can be written in the
form
\begin{equation} \label{advec_ready_L}
  v_{\tau} = \tilde L v, \qquad \tilde L = \tilde \Gamma :\nabla\nabla + \tilde \delta \cdot\nabla\,.
\end{equation}
Averaging this equation immediately yields at the leading order
\begin{equation}\label{av_eqn}
V_{\tau} = \langle \tilde L \rangle V\,.
\end{equation}
Here, and in what follows, $\langle...\rangle$ denotes averaging over one period. 
The averaged equation is then
\begin{equation}
V_{\tau} = \left(\langle \tilde \Gamma\rangle :\nabla\nabla + \langle \tilde \delta\rangle \cdot\nabla\right)V\,.
\end{equation}
Thus, the time-dependent coefficient are simply replaced by their time
averages. The leading order of the generating function $\phi_1$ is found as
\begin{equation}
\phi_1 = L_1\,V \equiv \left( \int_0^{\tau}\tilde L-\langle \tilde L \rangle \right)V\,.
\end{equation}
Introducing $\Gamma_1$ and $\delta_1$ as
\begin{equation} \label{int_functions}
\frac{d\Gamma_1}{d\tau} = \tilde \Gamma - \langle \tilde \Gamma \rangle, \qquad
\frac{d\delta_1}{d\tau} = \tilde \delta - \langle \tilde \delta \rangle
\end{equation}
we can write $L_1$ explicitly as
\begin{equation}
L_1 =   \Gamma_1 :\nabla\nabla + \delta_1 \cdot\nabla\,.
\end{equation}
Higher order corrections can be calculated in an elegant way 
using the Campbell-Baker-Hausdorff formulae. For the second order term
in the expansion of $Y$, for example, we find
\begin{eqnarray}
Y_1 &=& \frac{1}{2}\langle[L_1V,L_{1\tau}V]_L\rangle + [\langle L_1 \rangle V, \langle \tilde L \rangle V]_L \\ \nonumber
    &=& \left(\langle \tilde L L_1\rangle - \langle L_1 \rangle \langle \tilde L \rangle\right)V
\end{eqnarray}
where the last equality follows after using the definition of the Lie
commutator and integration by parts. Collecting first and second order,
we obtain as averaged equation for $V$ 
\begin{equation} \label{average_2nd_order}
V_{\tau} = \left(\langle \tilde L \rangle + \epsilon \left(\langle \tilde L L_1\rangle - \langle L_1 \rangle \langle \tilde L \rangle\right)\right)V
\end{equation}

\section{An averaging theorem for parabolic differential equations}

In the previous section, we applied a technique based on Lie
transforms to average the equation (\ref{advec_ready}), and we arrived
at the equation (\ref{av_eqn}).  Here, we state and prove rigorously a
theorem on averaging of parabolic partial differential equations which
is due to Krol \cite{krol:1991}. We assume that the differential
operators in (\ref{advec_ready}) are given by
\[
\tilde\Gamma =\epsilon[ a_{ij}(x, y,t)]_{i,j=1}^2
\]
and 
\[
\tilde\delta =\epsilon( b_1(t,x,y),  b_2(t,x,y)), 
\]
where $a_{i,j}, b_i\in C^{\infty}(\overline{R^2\times [0, \infty)})$,
and $[a_{ij}]$ is symmetric and uniformly positive definite, i.e,
there exists $\theta>0$ such that for all $\xi=(\xi_1,\xi_2)\in R^2$
and $(x,y,t)\in R^2\times [0,\infty)$ we have
\[
\sum_{i,j=1}^2a_{ij}(x,y,t)\xi_i\xi_j\geq\theta |\xi|^2. 
\] 
In the following, let $\tau_0=\mathcal{O}(1/\epsilon)$, and let
$\|\cdot\|_\infty$ denote the usual supremum norm on either $R^2$ or
$R^2\times[0,\tau_0]$, depending on the context.
\begin{theorem}
Let $v$ and $V$ be solutions to the Cauchy problems $v(0)=V(0)=v_0\in C^{\infty}(\overline{R^2})$ for the equations (\ref{advec_ready}) and (\ref{av_eqn}), respectively.  Then 
\[
\|v-V\|_\infty=\mathcal{O}(\epsilon).
\]
\end{theorem} 
{\bf Proof:}
First note that the existence and the uniqueness of bounded solutions
$v$, $V$ on $C^2(R^2\times[0,\tau_0])$ is well established (see
\cite{friedman:1964}).  Also, since (\ref{av_eqn}) is autonomous, the
derivatives $V_{\bf \alpha}$ of $V$ also satisfy an autonomous
parabolic differential equation with the same second order
differential operator, however with different smooth and bounded first
and zero order coefficients:
\[
V_{{\bf \alpha}\tau} =\left(\langle \tilde \Gamma\rangle :\nabla\nabla + \epsilon (b_1({\bf \alpha})(x,y),  b_2({\bf \alpha})(x,y)) \cdot\nabla + \epsilon f({\bf \alpha})(x,y) \right)V_{\bf \alpha}\,.
\]
The Phragm\`en-Lindel\"of principle for parabolic partial differential equations implies that 
\[
\|V_\alpha\|_\infty\leq \|(v_0)_\alpha\|_\infty  + \tau_0\epsilon \|f(\alpha)\|_\infty=\mathcal{O}(1)\,.
\]
Let us now define a near-identity transformation 
\[
\hat V(x,y,\tau)=V(x,y,\tau)+ \left[\int_0^\tau (\tilde L(s)-\langle\tilde L\rangle)\ ds\right] V(x,y,\tau). 
\]
Since the integrand is $T$-periodic with zero average, the equation reads actually 
\[
\hat V(x,y,\tau)=V(x,y,\tau)+ \left[\int_{[\tau/T]T}^\tau (\tilde L(s)-\langle\tilde L\rangle)\ ds\right] V(x,y,\tau), 
\]
and it is an easy observation that $\|\hat V-V\|_\infty =\mathcal{O}(\epsilon)$.  

On the other hand, one easily verifies that  $\hat V$ satisfies the equation 
\[
\hat V_\tau=\tilde L(\tau) \hat V+ \tilde M(\tau)V, 
\]
where $\tilde M(\tau)=\int_0^\tau (\tilde L(s)-\langle\tilde L\rangle)\langle\tilde
L\rangle-\tilde L(\tau)(\tilde L(s)-\langle\tilde L\rangle)\ ds$, and the initial-value
condition $\hat V(0)=v_0$. Notice that $\tilde M$ is a $T$-periodic
fourth order operator with smooth bounded coefficients of order
$\mathcal{O}(\epsilon^2)$.  Consequently, the difference $\hat V-v$
satisfies the equation
\[
(\hat V-v)_\tau=\tilde L(\tau)(\hat V-v) + \tilde M(\tau) V,
\]
and the initial condition $(\hat V-v)(0)=0$.  By the
Phragm\`en-Lindel\"of principle for parabolic partial differential
equations, we conclude $\|\hat V-v\|_\infty\leq \|\tilde M(\cdot)
V\|_\infty\tau_0=\mathcal{O}(\epsilon)$. This concludes the proof.

We remark that similar techniques can be applied to solutions of the second order Lie-averaged equations
leading to ${\mathcal{O}}(\epsilon^2)$ error estimates.

\section{Regularized vortical flow field}

To illustrate the theory and to give a comparison to numerical simulations, we
consider the particular case of a regularized vortical flow field whose stream
function is given by
\begin{equation}
\Psi(t,x,y) = \ln\left(\sqrt{a^2+x^2+y^2}\right)\,f(t) \;.
\end{equation}
This flow represents perhaps the simplest example for studying the interplay between diffusion
and nonlinear, time-periodic advection.

Since $r^2=x^2+y^2$ is a constant of motion, the unperturbed stream lines are
given in Cartesian coordinates as
\begin{eqnarray*}
x(t) &=& \cos\left(\omega(r)F(t)\right)x_0 + \sin\left(\omega(r)F(t)\right)y_0 \\
y(t) &=& -\sin\left(\omega(r)F(t)\right)x_0 + \cos\left(\omega(r)F(t)\right)y_0
\end{eqnarray*}
where
\begin{equation}
\omega(r) = \frac{1}{a^2+r^2}\,.
\end{equation}
Obviously, we can take as a canonical transform to action-angle variables the 
usual transformation to polar coordinates $(x,y)\rightarrow (r,\theta)$ and
in these coordinates, the advection-diffusion equation (\ref{advec}) is 
written for this particular flow field as
\begin{equation}
c_t - \omega(r)f(t)c_{\theta} = \epsilon\left(\frac{1}{r}c_r + c_{rr}+\frac{1}{r^2}c_{\theta\theta}\right)=\epsilon\Delta\,c\,.
\end{equation}
The transformed equation (\ref{advec_ready}) becomes (using $t$ instead of the rescaled $\tau$)
\begin{equation}
v_t = \epsilon\left(\Delta v + F\left(\left(\frac{\omega'}{r}+\omega''\right)v_{\tilde \theta}+2\omega'v_{\tilde \theta r}\right)+F^2(\omega')^2v_{\tilde\theta\tilde\theta}\right)
\end{equation}
%
%Introducing a multi-scale expansion
%
%\begin{equation}
%v(t,r,\tilde \theta) = V(t_1,t_2,...,r,\tilde\theta)+\epsilon V^{(1)}(t_0,t_1,t_2,...,r,\tilde\theta)+...
%\end{equation}
%
%with $t_k=\epsilon^k t$, 
%
Using the previous results, we obtain at leading order 
\begin{equation} \label{advec_vortex_av}
V_{t} = \epsilon\left(\Delta\,V + \langle F \rangle \left(\left(\frac{\omega'}{r}+\omega''\right) V_{\tilde\theta}
    +2\omega' V_{\tilde\theta r}\right) + \langle F^2\rangle(\omega')^2 V_{\tilde\theta\tilde\theta}\right)
\end{equation}

As shown in the Appendix, the leading order contributions in Cartesian
coordinates produce a time independent advection field with spatially
dependent rotation and source-like terms. The averaged diffusivity
tensor is usually full and symmetry breaking. The relative importance
of the symmetry breaking terms depends upon the explicit form of the
time dependence through the ratio of $\langle F\rangle $ and $\langle
F^2\rangle$.

For this flow-field, it is not difficult to compute corrections at second order as well.
In order to make our notation more efficient, we introduce the two 
operators ${\mathcal{L}}_1$ and ${\mathcal{L}}_2$
\begin{eqnarray}
{\mathcal{L}}_1 &\equiv& \left(\frac{\omega'}{r}+\omega''\right)\partial_{\tilde\theta}+ 2\omega'\partial_{\tilde\theta}\partial_{r} \\
{\mathcal{L}}_2 &\equiv& (\omega')^2\partial_{\tilde\theta}\partial_{\tilde\theta}
\end{eqnarray}
and introduce $F_1\equiv F$ and $F_2\equiv F^2$. The first order
averaged equation (\ref{advec_vortex_av}) becomes then
\begin{displaymath}
V_{t} =\epsilon\left( \Delta\,V + \langle F_1 \rangle {\mathcal{L}}_1 V +  \langle F_2 \rangle {\mathcal{L}}_2 V \right)
\end{displaymath}
%
%For the first order correction $V^{(1)}$ we find
%
%\begin{equation}
%V^{(1)}(t_0,t_1,t_2,...,r,\tilde\theta) = (G_1{\mathcal{L}}_1+G_2{\mathcal{L}}_2)V
%\end{equation}
%
Applying now (\ref{int_functions}), we introduce the functions $G_1$
and $G_2$ that can be found explicitly from $F_1$ and $F_2$ as
\begin{equation}
G_j(t_0) = \int_0^{t_0}(F_j(\tau)-\langle F_j \rangle)d\tau, \qquad j=1,2 \;\;.
\end{equation}
At second order in $\epsilon$ the equation is
\begin{eqnarray}
  V_{t} &=& \epsilon\left( \Delta\,V + \langle F_1 \rangle {\mathcal{L}}_1 V 
    +  \langle F_2 \rangle {\mathcal{L}}_2 V \right)
  +\epsilon^2 \left([\Delta,\langle G_1 \rangle {\mathcal{L}}_1]V \right. 
  + [\Delta,\langle G_2 \rangle {\mathcal{L}}_2]V \nonumber \\
  && + (\langle F_1G_1 \rangle -\langle F_1 \rangle \langle G_1 \rangle){\mathcal{L}}_1^2V
  + (\langle F_2G_2 \rangle -\langle F_2 \rangle \langle G_2 \rangle){\mathcal{L}}_2^2V \nonumber \\
  && + (\langle F_1G_2 \rangle -\langle F_2 \rangle \langle G_1 \rangle){\mathcal{L}}_1{\mathcal{L}}_2 V
  + \left. (\langle F_2G_1 \rangle -\langle F_1 \rangle \langle G_2 \rangle){\mathcal{L}}_2{\mathcal{L}}_1 V\right)
\label{second_order}
\end{eqnarray}
where $[A,B]\equiv AB-BA$ denotes the usual commutator.

\section{Numerical Simulations}

We can now integrate both the
original problem (\ref{advec}) and the averaged equation in first
order (\ref{advec_vortex_av}) numerically and compare their
solutions. In our numerical simulations, we use a standard
Adams-Bashforth-Moulton Method in Fourier space, the Fourier
transformations are done using FFTW.  We work back in Cartesian space and
for this purpose, we compare solutions at
Poincar\'e-sections where $F$ is zero, $\theta = \tilde
\theta$ and (\ref{advec_vortex_av}) is written in Cartesian
coordinates. 
The explicit form of (\ref{advec_vortex_av}) in Cartesian coordinates is given 
in the appendix. 
Throughout, we consider a single initial condition of the form
\begin{equation}
c^{(s)}(x,y)= x\,\exp(-b\,r^2)
\end{equation}
where the constants $a$ and $b$ were chosen as $a=1.0$ and $b=4.0$.

\begin{figure}[htb]
\centering
\begin{minipage}{1.0\textwidth}
\begin{center}
\includegraphics[width=0.489\textwidth, angle=0]{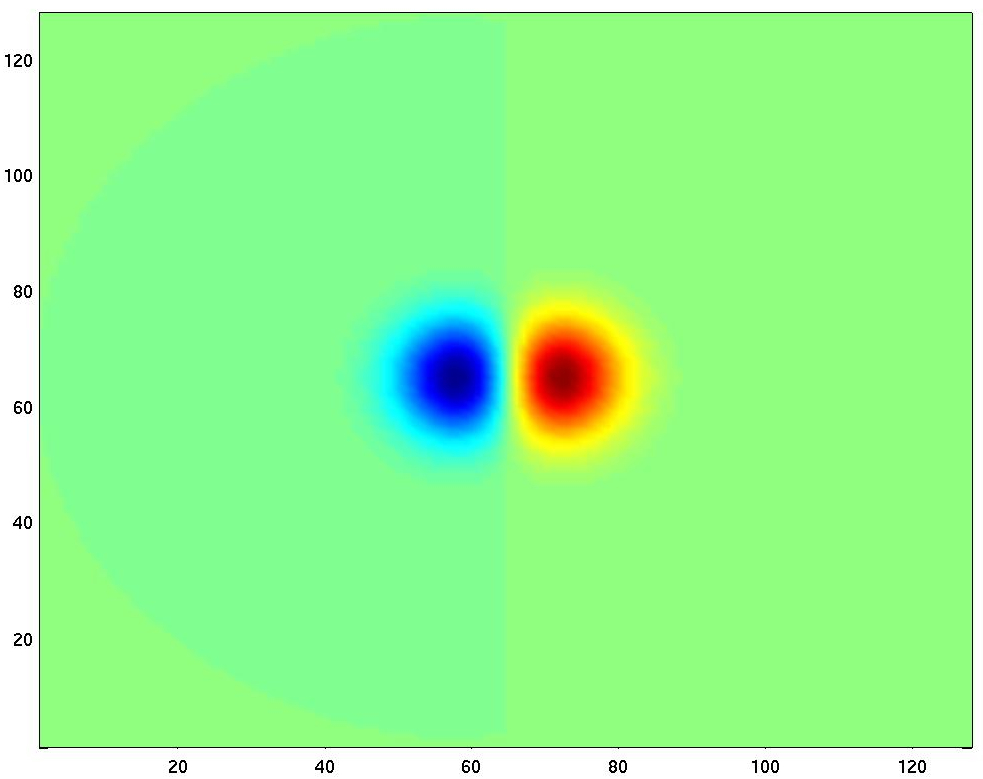}
\hfill
\includegraphics[width=0.489\textwidth, angle=0]{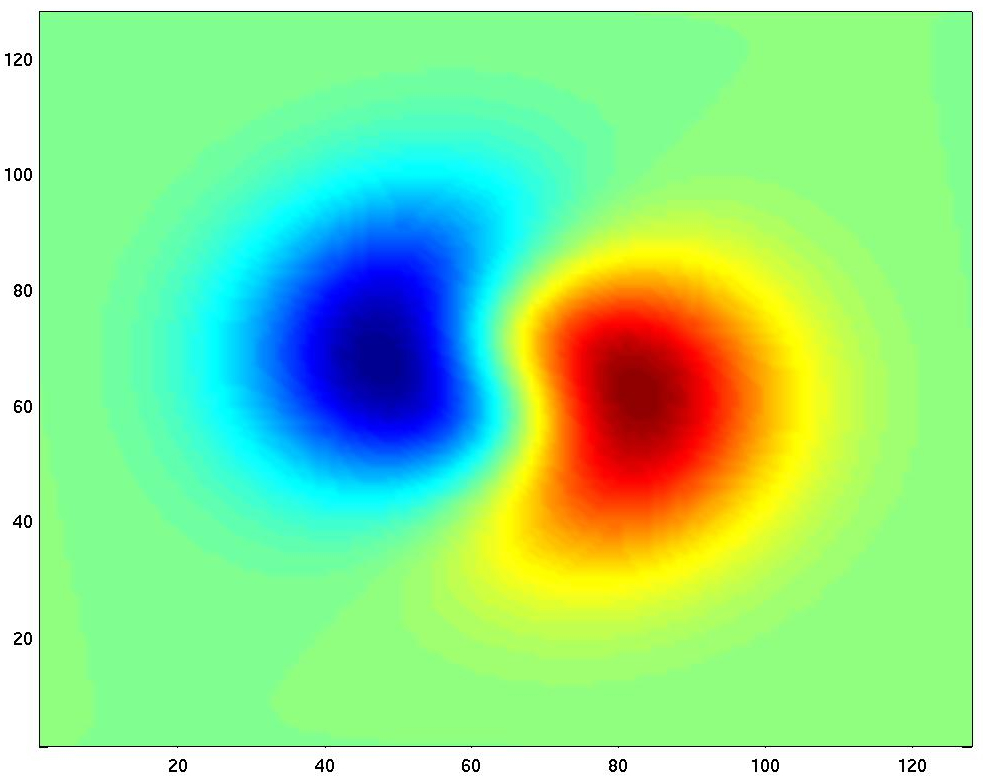}
\caption{{\small Evolution of tracers in the time-periodically modulated vortical flow field.
The figure on the left shows the initial condition, the figure on the right shows the result 
of the tracer motion after 10 periods.}}
\label{fig:advec}
\end{center}
\end{minipage}
\end{figure}

As shown in (\ref{advec_vortex_av}), the form of the averaging depends explicitly on the nature of $f(t)$.
For the particular case of $f(t)=\sin(t)$, we find $\langle F \rangle = 1$ 
and $\langle F^2 \rangle = 3/2$. 
Fig. \ref{fig:advec} shows both the
initial condition and the evolution after ten periods for this choice of $f(t)$.

\begin{figure}[htb]
\centering
\begin{minipage}{1.0\textwidth}
\begin{center}
\includegraphics[width=0.485\textwidth, angle=0]{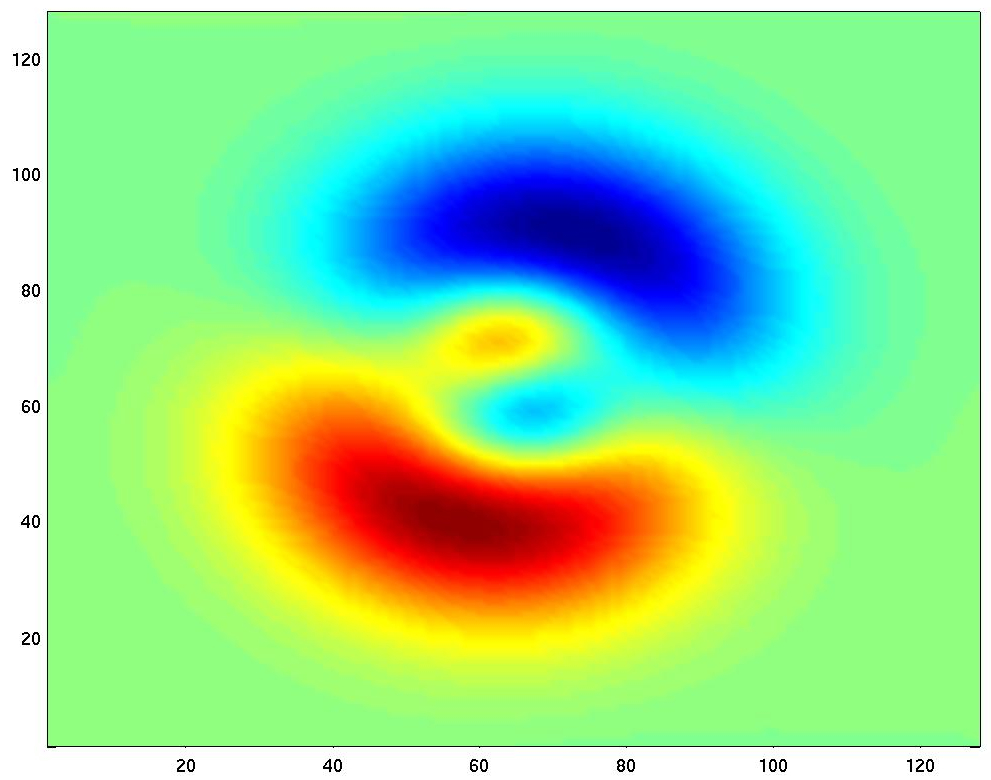}
\hfill
\includegraphics[width=0.485\textwidth, angle=0]{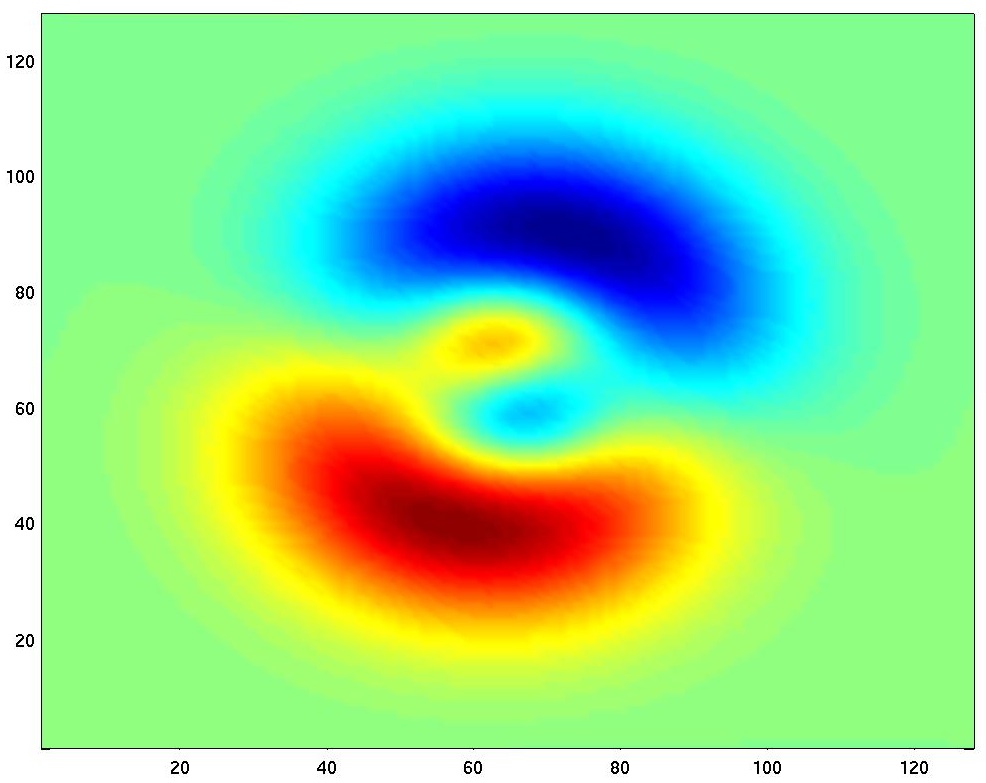}
\caption{{\small Comparison of the prediction of the averaged equation
    (\ref{advec_vortex_av}) to the evolution of the time-dependent
    equation (\ref{advec}). We plot the difference between the purely
    viscous solution and the solution incorporating the effects of the
    time-dependent advection field. The figure on the left shows
    $c-c^{\mathrm{vis}}$ where $c$ is the solution of
    (\ref{advec}). The figure on the right shows
    $c^{(\mathrm{av})}-c^{(\mathrm{vis})}$, where $c^{(\mathrm{av})}$
    has been found using the averaged equation
    (\ref{advec_vortex_av}). The averaged equation is obviously able
    to capture the leading order impact of the time-dependent velocity
    field.}}
\label{fig:advec_comp}
\end{center}
\end{minipage}
\end{figure}

In the absence of the time-dependent
advection field, the tracer field will obey the purely diffusive equation
\begin{equation} \label{diff_eq}
c^{(\mathrm{vis})}_t = \epsilon \Delta c^{({\mathrm{vis}})}, \qquad c^{(\mathrm{vis})}(0,\xi)=c^{(s)}(\xi)
\end{equation}
and the diffusion will simply spread the initial distribution out and
preserve its symmetries. In the presence of the time-dependent
vortical field, however, the particles will move back and forth within
one period and the interplay of the time-dependent trajectories with
the diffusion will give rise to a breaking of the symmetry of the
initial distribution resulting in a ``twist'' that can be clearly seen
in the right figure of Fig. \ref{fig:advec}. In order to determine how well
the averaged equation (\ref{advec_vortex_av}) captures the
differences between the purely diffusive case and the case with both
time-dependent advection and diffusion, 
the difference of the purely viscous case and the solution
of (\ref{advec}), hence $c-c^{(\mathrm{vis})}$ and (b) the difference of
the purely viscous case and the approximation $c^{(\mathrm{av})}$
constructed using (\ref{advec_vortex_av}), hence
$c^{(\mathrm{av})}-c^{(\mathrm{vis})}$ are plotted in Fig. \ref{fig:advec_comp}.
The first order approximation accurately captures the overall dynamics of the full,
time-dependent problem.

\begin{figure}[htb]
\centering
\begin{minipage}{1.0\textwidth}
\begin{center}
\includegraphics[width=0.5\textwidth, angle=90]{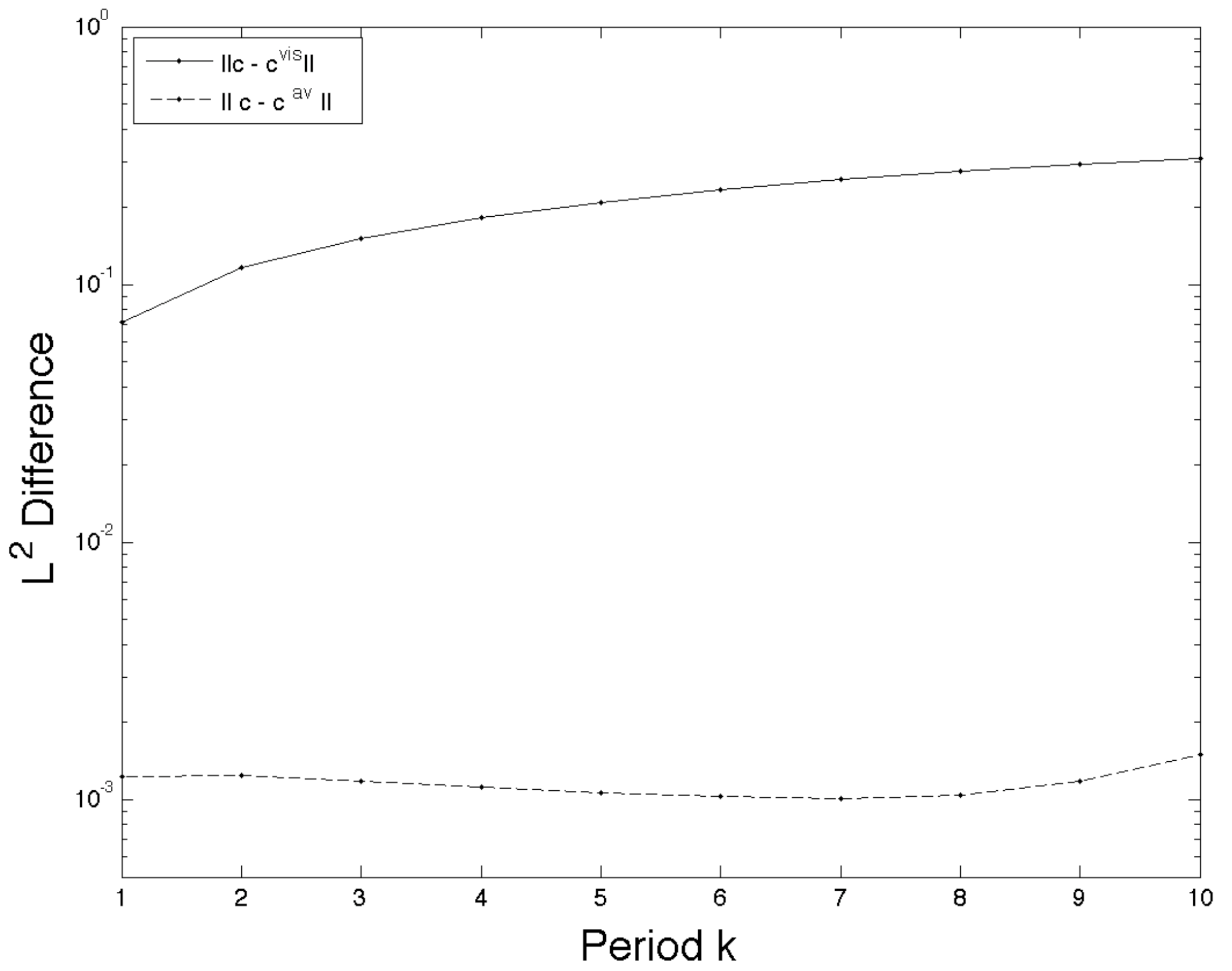} \includegraphics[width=0.5\textwidth, angle=90]{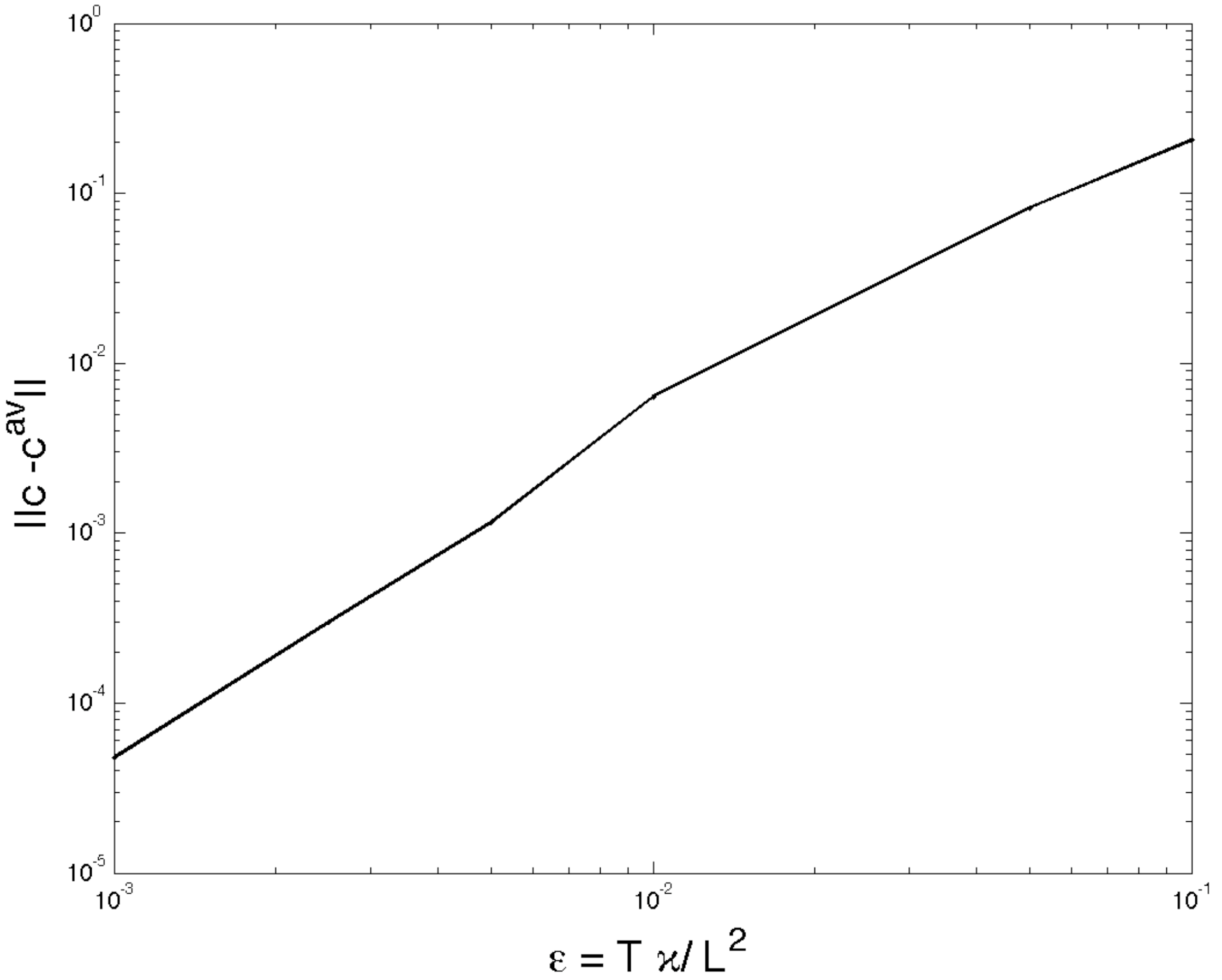}
\caption{{\small (a) Difference norms for the case $f(t) = \sin(t)$ and $\epsilon = 0.005$. (b) Average error between the first order averaged equation and the full solution versus $\epsilon$ for $f(t) = \sin(t)$. The 
error scales like $\epsilon^{1.8}$. }}
\label{fig:comp-L2}
\end{center}
\end{minipage}
\end{figure}

To quantify the accuracy of the approximation, we 
consider the time-evolution of the canonical
$L^2$-norm of the differences by defining
\begin{equation}
\|c-c^{(av)}\| = \left(\frac{\int_{\mathbb{R}^2}|c-c^{(av)}|^2\;dx\,dy}{\int_{\mathbb{R}^2}|c|^2\;dx\,dy}\right)^{1/2}
\end{equation}
Figure \ref{fig:comp-L2}a shows that this error is small and approximately constant
over the first 10 periods whereas the corresponding error between $c$ and
$c^{(\mathrm{vis})}$ is comparatively large and growing exponentially in time.
Figure \ref{fig:comp-L2}b indicates that, at least for the case where $f(t) = \sin(t)$, the solutions
of the averaged equation converge to those of the full equation faster than $\epsilon$. 

\begin{figure}[htb]
\centering
\begin{minipage}{1.0\textwidth}
\begin{center}
\includegraphics[width=0.85\textwidth, angle=0]{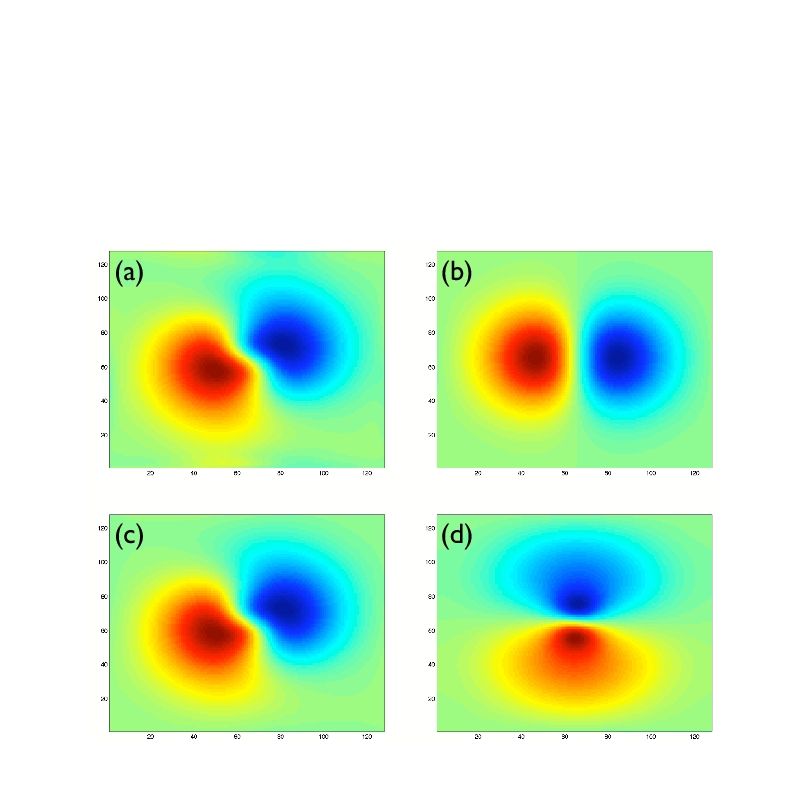} 
\caption{{\small Solutions for the case $f(t) = \cos(t)$ and $\epsilon = 0.010$ after 10 periods. 
(a) Full solution, (b) first order averaged solution, (c) second order averaged solution, (d) difference between first and second order averaged. }}
\label{fig:cosine_comp}
\end{center}
\end{minipage}
\end{figure}

\begin{figure}[htb]
\centering
\begin{minipage}{1.0\textwidth}
\begin{center}
\includegraphics[width=0.5\textwidth, angle=90]{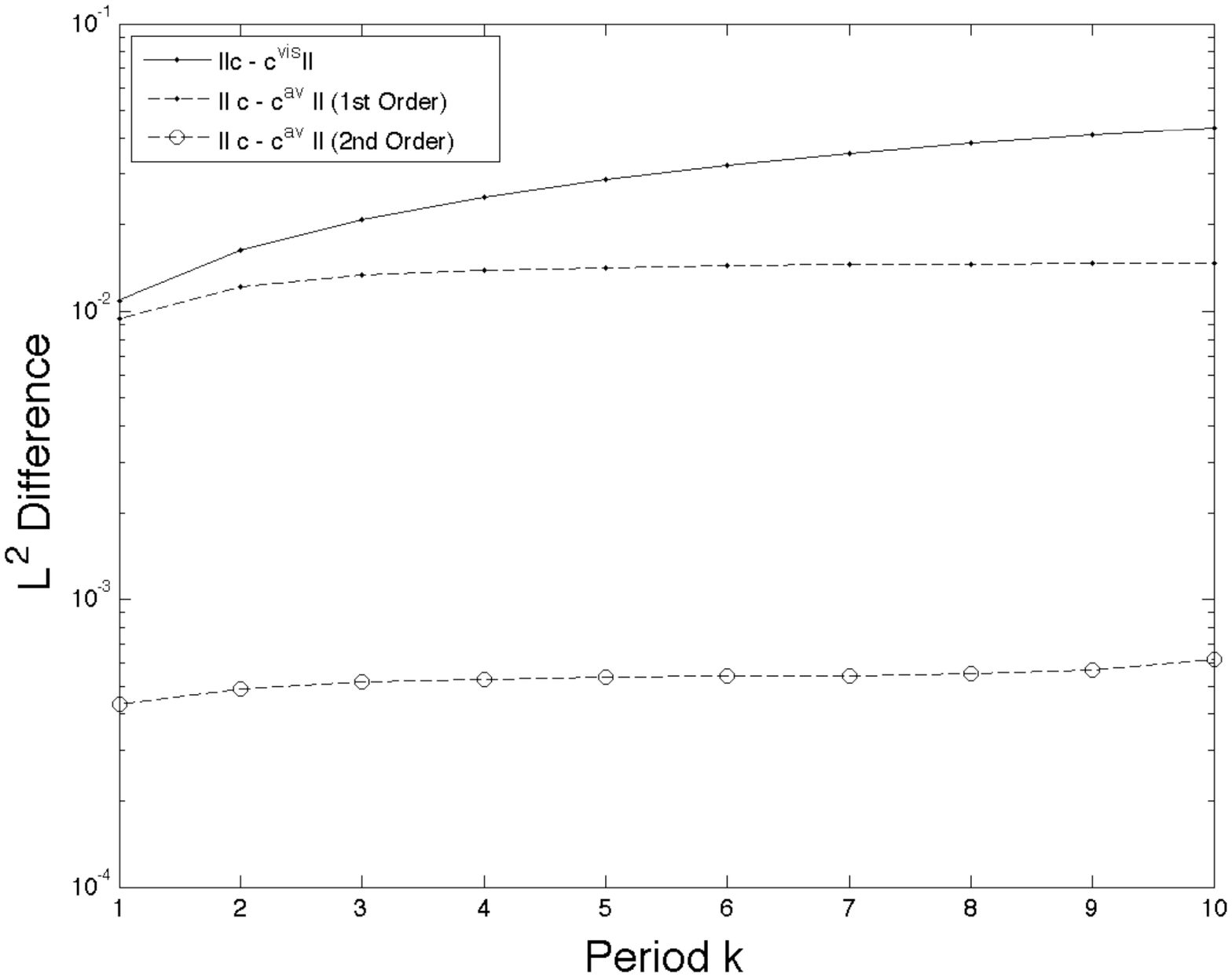} \includegraphics[width=0.5\textwidth, angle=90]{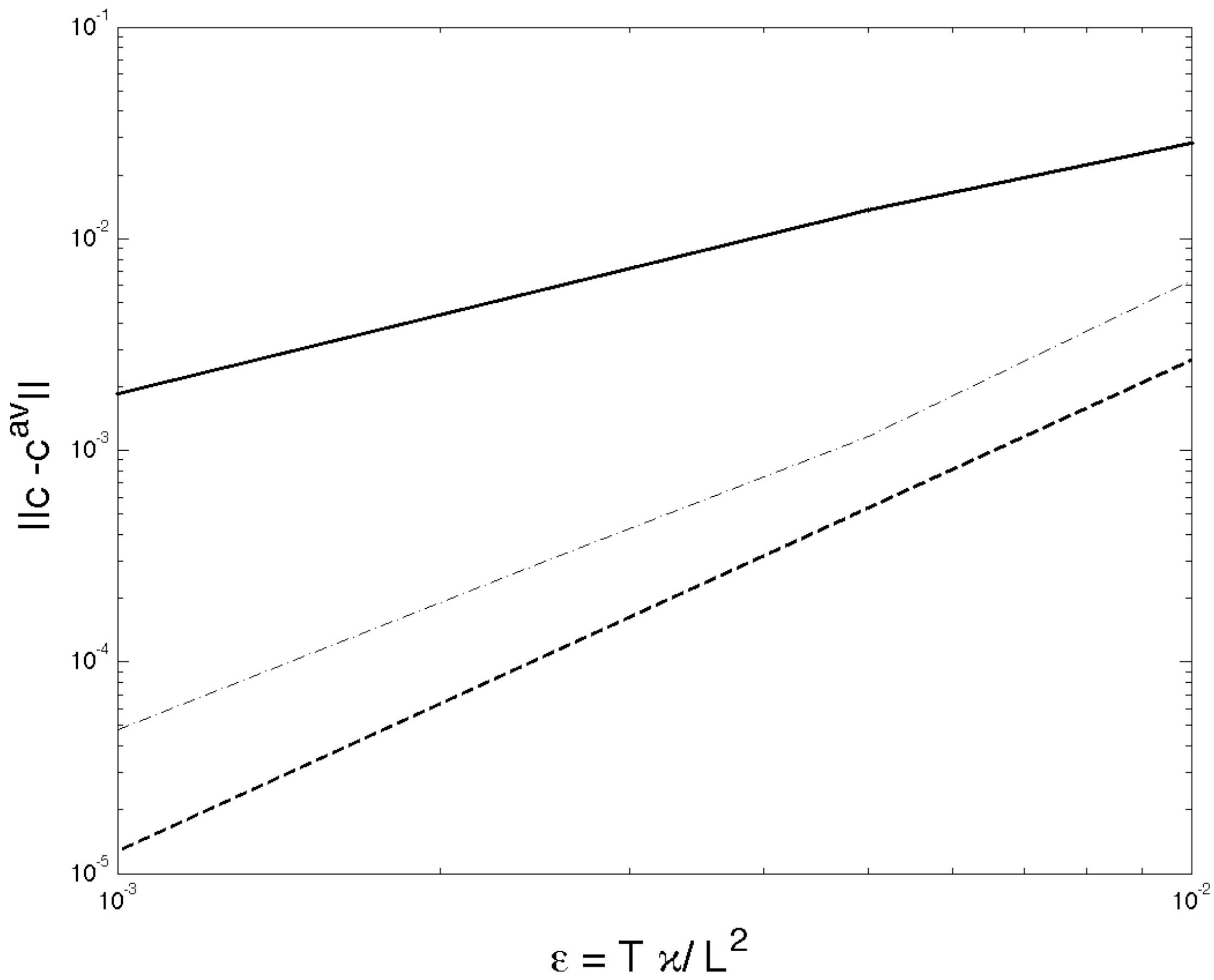}
\caption{{\small (a) Difference norms for the case $f(t) = \cos(t)$ and $\epsilon = 0.005$ for both first and second order averaging. (b) Average error between the averaged equation and the full solution versus $\epsilon$ for $f(t) = \cos(t)$. The first order solution is shown by the solid curve, the second order by the dashed curve. For comparison, the first order results for $f(t) = \sin(t)$ are shown in the light dot-dashed line. }}
\label{fig:comp-L2-cos}
\end{center}
\end{minipage}
\end{figure}

We examine the role of $f(t)$ in the averaged dynamic by setting  $f(t) = \cos(t)$. 
This choice implies that $\langle F\rangle = 0$, leading to near degeneracy of the first order corrections. 
In this case, the time independent advection terms produced by the averaging procedure  do not 
contribute to 'twisting' the scalar evolution. The symmetry breaking terms in the averaged diffusivity tensor are identically zero at
first order. The effect of this near-degeneracy for $\langle F \rangle = 0$ is clearly seen in the comparison of 
panels (a) and (b) in Fig. \ref{fig:cosine_comp}. 
A comparison of the difference norms, shown in Fig. \ref{fig:comp-L2-cos}(a), indicates that the
first order averaged equation is only a marginal improvement on the purely viscous solution at short
times. 
However, as shown in Fig. \ref{fig:comp-L2-cos}(b), solutions to the first order averaged equations 
continue to converge to the true solution with decreasing $\epsilon$ although the convergence rate, 
$\sim \epsilon^{1.2}$, is considerably slower than that observed when $f(t)=\sin(t)$.

For $\langle F \rangle =0$, second order contributions are clearly
important. Referring back to (\ref{second_order}), this situation also
leads to a relatively simple form for the second order expression. The
coefficient in front of ${\mathcal{L}}_1$ vanishes, and in this
particular case for $f(t)=\cos(t)$, (\ref{second_order}) simplifies to
\begin{equation}
V_{t} = \epsilon \left(\Delta\,V+ \frac{1}{2} {\mathcal{L}}_2\,V\right) + \epsilon^2\left([\Delta,{\mathcal{L}}_1]V - \frac{1}{2}[{\mathcal{L}}_1,{\mathcal{L}}_2]V\right)
\end{equation}
Evaluating this equation explicitly we find
\begin{eqnarray}
V_{t} &=& \epsilon \left(\Delta\,V + (\omega')^2V_{\tilde\theta\tilde\theta}\right)+\epsilon^2\left(\left(\frac{\omega'}{r^3}-\frac{\omega''}{r^2}+2\frac{\omega'''}{r}+\omega^{(iv)}\right)V_{\tilde\theta}\right.\nonumber \\
&&
+4\left(\frac{\omega''}{r}+\omega'''\right)V_{{\tilde\theta r}} +\left.\left(4\frac{\omega'}{r^3}-2(\omega')^2\omega''\right)V_{{\tilde\theta}{\tilde\theta}{\tilde\theta}}+4\omega''V_{\tilde\theta r r}\right) \label{eq:av_cos_polar}
\end{eqnarray}

As shown in the Appendix, this leads to ${\mathcal{O}}(\epsilon)$ symmetry breaking contributions to both
the average advection terms and diffusivity tensor as well as contributions in the form of higher,
third order, spatial derivatives. Numerically, such terms are easily computed spectrally. An example of a 
second order solution is shown in panel (c) of Fig. \ref{fig:cosine_comp}. Obviously, the second order 
solution is a clear improvement on the first order approximation and the difference between the two, 
shown in panel (d), demonstrates the restoration of advective twist at higher order. 
Figure \ref{fig:comp-L2-cos} indicates both the increase in accuracy and the expected increase in 
convergence rate for the second order approximation. 

\section{Discussion}

We have proposed a scheme for formally transforming the advection
diffusion equation, in the limit of small diffusivity, into a form
suitable for averaging. We have given an explicit means of averaging
the transformed equation and proven the convergence of solutions of
the averaged, time-independent approximation to the full
dynamics. Throughout, however, we deal only with a restricted class of
advecting fields, namely those which are zero-mean and possess
time-periodic stream lines. While such fields are inherently
integrable, and hence explicitly non-chaotic, the results presented
are of interest in the study of 'strange eigenmodes' of the advection
diffusion equation in the Batchelor regime. First, the emergence of
strange eigenmodes is independent of the integrability or
non-integrability of the underlying flow \cite{Camassa:2007}, and
indeed, the simple example considered here produces non-trivial
periodic patterns. Secondly, the results shown for even these
extremely simple cases point to the delicate relationship between the
non-linearity of the conservative advection operator and small
diffusivity. The analysis points to fundamental differences between
the dynamics of continuous time flows and discrete time maps. Indeed,
for the periodically modulated vortex considered, the averaged
dynamics depends strongly on the phase of the single frequency
periodic modulation, a fact completely lost when considering the
Poincar{\'e} map of the flow which is simply the identity.

In the case of a mean-free advection field with periodic stream lines,
the transform to action-angle variables of the Lagrangian flow was
found to be the appropriate transformation for deriving accurate
time-averaged dynamics in the limit of small diffusivity. Finding
transformations with similar properties for other classes of advection
fields will likely provide a means for understanding the interplay
between advection and small diffusion.

\vspace{1.in}

\noindent \textbf{Acknowledgment}
ACP, TS and JV were supported, in part, by a grant from the City University of New York PSC-CUNY Research Award Program. The authors gratefully acknowledge the support of the CUNY High Performance Computing Facility and the Center for Interdisciplinary Applied Mathematics and Computational Sciences.  Also, JV was supported in part by the NSF grant DMS-0733126. 

\vspace{0.25in}

\appendix

\section{The averaged equation in Cartesian coordinates}

In order to convert (\ref{advec_vortex_av}) back to Cartesian coordinates,
we note first that it follows directly from the transformation rules to polar coordinates that
\begin{displaymath}
r\partial_r = x\partial_x+y\partial_y, \qquad \partial_{\theta} = x\partial_y - y\partial_x
\end{displaymath}
which, after straightforward calculations, yield the following transformation
rules for the higher operators occurring in (\ref{advec_vortex_av}):
\begin{eqnarray*}
\partial_{\theta}\partial_{\theta} &=& -x\partial_x - y\partial_y \\
                                && + y^2\partial_x\partial_x - 2xy \partial_x\partial_y + x^2\partial_y\partial_y \\
r\partial_{\theta}\partial_r &=& x\partial_y - y\partial_x \\
                            && -xy\partial_x\partial_x + (x^2-y^2)\partial_x\partial_y + xy\partial_y\partial_y
\end{eqnarray*}
Rescaling time as $\tau = \epsilon t$, we find for (\ref{advec_vortex_av}) then
\begin{equation}
V_{\tau} + (\Xi \cdot\nabla)V = K : \nabla\nabla V
\end{equation}
where the advection vector $\Xi$ and is given as
\begin{eqnarray*}
\Xi_1(x,y) &=& \langle F \rangle \left(\left(\frac{\omega'}{r}+\omega''\right)+2\frac{\omega'}{r}\right)y
                 +2\langle F^2 \rangle (\omega')^2 x \\
\Xi_2(x,y) &=& -\langle F \rangle \left(\left(\frac{\omega'}{r}+\omega''\right)+2\frac{\omega'}{r}\right)x
                 +2\langle F^2 \rangle (\omega')^2 y
\end{eqnarray*}
and the diffusion matrix $K$ becomes
\begin{eqnarray*}
K_{11} &=& -2\frac{\omega'}{r}xy \langle F \rangle +\langle F^2 \rangle (\omega')^2y^2 \\
K_{12} &=& K_{21} = \frac{\omega'}{r}\langle F \rangle (x^2-y^2)-\langle F^2 \rangle (\omega')^2xy \\
K_{22} &=& 2\frac{\omega'}{r}xy \langle F \rangle +\langle F^2 \rangle (\omega')^2x^2
\end{eqnarray*}
This together with the formulas
\begin{displaymath}
\omega' = \frac{-2r}{\left(a^2+r^2\right)^2}, \qquad \frac{\omega'}{r}+\omega'' = 4\frac{r^2-a^2}{(r^2+a^2)^3}
\end{displaymath}
gives directly the coefficients of (\ref{advec_vortex_av}) in
Cartesian coordinates. Note that all operators give a contribution to
the averaged advection field in Cartesian coordinates.

For the second-order, the corresponding terms are more complicated. The explicit transformations of 
the operators are given by
\begin{eqnarray*}
r^2\partial_{\theta}\partial_r^2 &=& 2xy\partial_y^2+2(x^2-y^2)\partial_x\partial_y-2xy\partial_x^2 \\ 
&& + xy^2\partial_y^3+(x^3-2xy^2)\partial_x^2\partial_y + (2x^2y-y^3)\partial_x\partial_y^2 -x^2y\partial_x^3\\
\partial_{\theta}^3 &=& -x\partial_y+y\partial_x \\
                    && + 3xy\partial_x^2 +3(y^2-x^2)\partial_x\partial_y - 3xy\partial_y^2 \\
                    && + x^3\partial_y^3 - 3x^2y\partial_x\partial_y^2 + 3xy^2 \partial_x^2\partial_y -y^3\partial_x^3
\end{eqnarray*}
Note in particular that the first term in the transformation of $\partial_{\theta}^3$ corresponds to 
$-\partial_{\theta}$, hence is one example of a term that will have a "twisting" effect on the 
solution and that can lead to symmetry breaking.
We can now rewrite (\ref{eq:av_cos_polar}) in Cartesian coordinates (rescaling time $\tau=\epsilon t$
as before) and obtain:
\begin{eqnarray}
V_{\tau} + \left(\Xi \cdot \nabla\right)V &=& K_{11}V_{xx}+2K_{12}V_{xy}+ K_{22}V_{yy} \nonumber \\ 
 && + M_{111}V_{xxx}+M_{112}V_{xxy}+M_{122}V_{xyy}+M_{222}V_{yyy}
\end{eqnarray}
where the coefficients of the advection field $\Xi$ are given by
\begin{eqnarray*}
\Xi_1 &=& f_5x+\epsilon\left((f_1-f_3)y+f_2\frac{y}{r}\right)\\
\Xi_2 &=&f_5y+\epsilon\left((f_3-f_1)x-f_2\frac{x}{r}\right),
\end{eqnarray*}
and the diffusion matrix $K$ results in
\begin{eqnarray*}
K_{11} &=& 1+f_5y^2 +\epsilon\left(3f_3xy-f_2\frac{xy}{r}-2f_4\frac{xy}{r^2}\right)  \\
K_{22} &=& 1+f_5x^2 +\epsilon\left(-3f_3xy+f_2\frac{xy}{r}+2f_4\frac{xy}{r^2}\right) \\
2K_{12}&=& -2f_5xy +\epsilon\left(f_2\frac{x^2-y^2}{r}+3f_3(y^2-x^2)+2f_4\frac{x^2-y^2}{r^2}\right)\,.
\end{eqnarray*}
As for the tensor $M$ yielding higher-order contributions, we find
\begin{eqnarray*}
M_{111}&=&-f_3y^3-f_4\frac{x^2y}{r^2} \\
M_{222}&=&f_3x^3+f_4\frac{xy^2}{r^2} \\
M_{112}&=&3f_3xy^2+f_4\frac{x^3-2xy^2}{r^2} \\
M_{122}&=&-3f_3x^2y-f_4\frac{y^3-2x^2y}{r^2}
\end{eqnarray*}
and the coefficients $f_k$ are explicitly written as
\begin{eqnarray*}
f_1 &=& \frac{\omega'}{r^3}-\frac{\omega''}{r^2}+2\frac{\omega'''}{r}+\omega^{(iv)} 
     = 64\frac{r^4-4a^2r^2+a^4}{(r^2+a^2)^5} \\
f_2 &=& 4\left(\frac{\omega''}{r}+\omega'''\right)
     = -\frac{8}{r}\cdot\frac{9r^4-14a^2r^2+a^4}{(r^2+a^2)^4} \\
f_3 &=& 4\frac{\omega'}{r^3}-2(\omega')^2\omega'' 
     = -\frac{8}{r^2(r^2+a^2)^2}+16\frac{a^2r^2-3r^4}{(r^2+a^2)^7} \\
f_4 &=& 4\omega'' = 4\cdot\frac{6r^2-2a^2}{(r^2+a^2)^3} \\
f_5 &=& \frac{1}{2}(\omega')^2 = \frac{2r^2}{(a^2+r^2)^4}\,.
\end{eqnarray*}

\bibliography{master}

\end{document}